\documentclass[prd,twocolumn,nofootinbib, preprintnumbers, superscriptaddress, floatfix, longbibliography]{revtex4-1}

\usepackage{amsmath,amssymb,slashed,braket,enumitem,soul}
\usepackage{graphicx}
\usepackage{epstopdf}
\usepackage{float}
\usepackage[colorlinks=true,
            linkcolor=blue,
            urlcolor=blue,
            citecolor=green,          
            bookmarks=true,
            bookmarksnumbered=true,
            breaklinks=true,
            pdfpagemode=FullScreen,
            pdfstartview=FitBH]{hyperref}
            
\setstcolor{red}

\DeclareUnicodeCharacter{02B9}{'}

\usepackage[normalem]{ulem}
\usepackage{cancel}
\usepackage{color}

\definecolor{Orange}{cmyk}{0,0.61,0.87,0}
\definecolor{JungleGreen}{cmyk}{0.99,0,0.52,0}
\definecolor{OliveGreen}{cmyk}{0.64,0,0.95,0.40}
\definecolor{Brown}{cmyk}{0,0.81,1,0.60}
\definecolor{RoyalBlue}{cmyk}{0.71,0.53,0,0.12}
\definecolor{Gray}{cmyk}{0,0,0,0.40}
\definecolor{LightPink}{cmyk}{0.0,0.25,0,0}
\definecolor{LLightPink}{cmyk}{0.0,0.10,0,0}
\definecolor{LightBlue}{cmyk}{0.25,0,0,0}
\definecolor{LightGray}{cmyk}{0,0,0,0.2}

\usepackage{xcolor}
\definecolor{gesfpurple}{rgb}{0.47,0.19,0.42}

\definecolor{gesflanse}{rgb}{0.00,0.50,0.50}

\definecolor{gesfblue}{rgb}{0.08,0.42,0.76}

\definecolor{gesfred}{rgb}{1,0,0}

\definecolor{gesfwhite}{rgb}{1,1,1}

\definecolor{gesfblack}{rgb}{0,0,0}

\newcommand{\geqn}[1]{Eq.\,\hypersetup{linkcolor=blue}(\ref{#1})\hypersetup{linkcolor=blue}}
\newcommand{\gfig}[1]{{\hypersetup{linkcolor=violet}Fig.\,\ref{#1}\hypersetup{linkcolor=blue}}}

\graphicspath{{figs/}}

\begin{document}

\title{Ultraheavy atomic dark matter freeze-Out through rearrangement}

\author{Yu-Cheng Qiu}
\email{ethanqiu@sjtu.edu.cn}
\affiliation{Tsung-Dao Lee Institute \& School of Physics and Astronomy, Shanghai Jiao Tong University, Shanghai
200240, China}
\affiliation{Key Laboratory for Particle Astrophysics and Cosmology (MOE) \& Shanghai Key Laboratory for Particle Physics and Cosmology, Shanghai Jiao Tong University, Shanghai 200240, China}
\author{Jie Sheng}
\email{shengjie04@sjtu.edu.cn}
\affiliation{Tsung-Dao Lee Institute \& School of Physics and Astronomy, Shanghai Jiao Tong University, Shanghai
200240, China}
\affiliation{Key Laboratory for Particle Astrophysics and Cosmology (MOE) \& Shanghai Key Laboratory for Particle Physics and Cosmology, Shanghai Jiao Tong University, Shanghai 200240, China}
\author{Liang Tan}
\email{tanliang@sjtu.edu.cn}
\affiliation{Tsung-Dao Lee Institute \& School of Physics and Astronomy, Shanghai Jiao Tong University, Shanghai
200240, China}
\affiliation{Key Laboratory for Particle Astrophysics and Cosmology (MOE) \& Shanghai Key Laboratory for Particle Physics and Cosmology, Shanghai Jiao Tong University, Shanghai 200240, China}
\author{Chuan-Yang Xing}
\email{chuan-yang-xing@sjtu.edu.cn}
\affiliation{Tsung-Dao Lee Institute \& School of Physics and Astronomy, Shanghai Jiao Tong University, Shanghai
200240, China}
\affiliation{Key Laboratory for Particle Astrophysics and Cosmology (MOE) \& Shanghai Key Laboratory for Particle Physics and Cosmology, Shanghai Jiao Tong University, Shanghai 200240, China}

\begin{abstract}

Atomic dark matter is usually considered to be produced asymmetrically in the early Universe.
In this work, we first propose that the 
symmetric atomic dark matter
can be thermally produced through the freeze-out mechanism. 
The dominant atom antiatom annihilation channel is the atomic rearrangement. It has a geometrical cross section
much larger than that of elementary fermions.
After the atomic formation, this annihilation process further depletes dark matter particles and finally freezes out.
To give the observed dark matter relic, the dark atoms are naturally ultraheavy, ranging from $10^6$ to $10^{10} \,\mathrm{GeV}$.

\end{abstract}

\maketitle

\section{Introduction}

More than 80\% of the matter in our Universe today is dark matter (DM)~\cite{Planck:2018vyg}. 
But the nature of DM is still a mystery that suggests new physics beyond the Standard Model (SM) of particle physics~\cite{Bertone:2004pz,Young:2016ala,Arbey:2021gdg}. 
Among all DM candidates, the weakly interacting massive particle (WIMP) scenario~\cite{Steigman:1984ac,Goodman:1984dc,Jungman:1995df,Arcadi:2017kky,Roszkowski:2017nbc} is widely recognized. 
Its production mechanism, the thermal freeze-out~\cite{Lee:1977ua,Kolb:1990vq}, is quite natural and attractive. The observed DM density is simply determined by the fundamental parameters, 
such as the DM mass and its coupling to the SM particles.

However, WIMP predictions have several conflicts with the observations, such 
as the small scale problems~\cite{Vogelsberger:2015gpr,Bullock:2017xww}.
Besides, 
the DM with a mass of WIMP scale has received strong constraints from direct detection experiments~\cite{Schumann:2019eaa,PandaX-4T:2021bab,Akerib:2022ort,LZ:2022lsv,XENON:2023cxc}.
Another appealing candidate,
atomic dark matter 
~\cite{1983AZh....60..632B,1986PhLB..174..151G,Kaplan:2009de,Kaplan:2011yj,Cline:2012is,Cline:2021itd},
appears as an extension of dark $U(1)$ gauge symmetry.
Since dark atom particles are naturally self-interacting, this atomic DM scenario can be utilized to solve the small scale problems
~\cite{Cyr-Racine:2012tfp,Cline:2013pca,Petraki:2014uza,Boddy:2016bbu,Agrawal:2017rvu,Tulin:2017ara,Bansal:2022qbi}.
In addition, atomic DM also has rich phenomenology in the direct detection~\cite{Kaplan:2011yj} and indirect detection~\cite{Pearce:2015zca}
experiments.

Being an analogy to the atoms in the SM sector, the atomic DM in the current literature is always designed as asymmetric. 
Its production is usually considered as
the out-of-equilibrium decay of right-handed neutrinos in the early Universe
~\cite{Kaplan:2011yj,Choquette:2015mca}
as an extension of leptogenesis\cite{Fukugita:1986hr}.
The assumption that the production of dark atoms occurs through thermal freeze-out is compelling since it reduces model complexity and provides a rich
phenomenology while leaving other possibilities unexplored.

\begin{figure}[t]
    \centering
    \includegraphics[width=7cm]{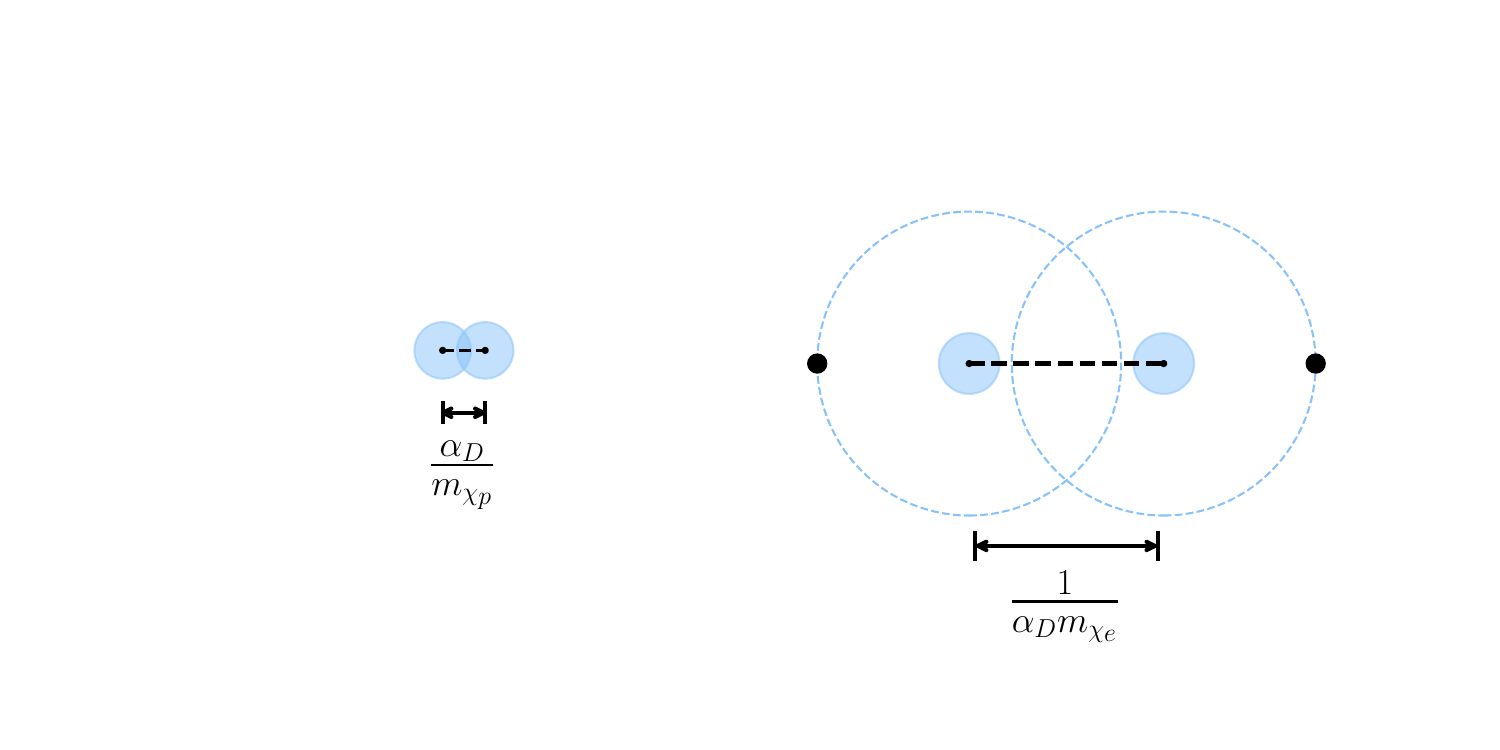}
    \caption{Illustration of the annihilation cross sections of elementary dark proton, $\chi_{p}$, and atomic bound state, $\chi_A$ (not to scale).
    The elementary fermions $\chi_{p}$ annihilate when their distance is smaller than $ \sim \alpha_{D}/m_{\chi_p}$, while the dark atoms would rearrange within a distance $\sim 1/\alpha_{D} m_{\chi_e}$. 
    The latter process is hugely enhanced for $\alpha_{D} < 1$ and $m_{\chi_p} \gg m_{\chi_e}$.}
    \label{fig:CrossSection}
\end{figure}

In this work, we first point out that the DM can be symmetric atomic states, which are thermally 
produced through 
freeze-out to give the 
observed relic.
The dark sector contains 
a heavier fermion 
$\chi_p$ and its lighter
partner $\chi_e$ with opposite dark $U(1)_X$ charges, as 
well as the same amount of their antiparticles.
When the Universe cools down, the dark fermions 
begin to form atomic bound states.
The dark atom mass
is dominated by the 
heavier fermion 
$\chi_p$, while its radius 
is determined by the 
lighter one $\chi_e$ \cite{sakurai2014modern}.
After formation, dark atom and antiatoms annihilate
by experiencing an intermediate state of dark
positronium and protonium.
This is the so-called
atomic rearrangement.
Its cross section is geometrical and proportional to the atomic size, which is much larger than 
that of a single fermion,
as shown in the cartoon in~\gfig{fig:CrossSection}.~\footnote{The self-destructing DM has a similar nature that it annihilates after rearrangement\cite{Geller:2020zhq,Grossman:2017qzw}.}
Thus, the DM number density is further depleted 
and the eventual freeze-out is determined by the atomic rearrangement.
As an important result, the symmetric atomic DM scenario naturally produces ultraheavy DM beyond the unitarity bound~\cite{Griest:1989wd,vonHarling:2014kha,Smirnov:2019ngs}.~\footnote{
Several exquisite mechanisms, such as Refs.~\cite{Harigaya:2016nlg,Berlin:2017ife,Kim:2019udq,Kramer:2020sbb,Frumkin:2022ror}, can also produce ultraheavy or hyperheavy DM thermally.}

After introducing such a scenario and its freeze-out history, we
explore the viable parameter space.
Finally, possible signatures and constraints of the symmetric dark atom scenario are discussed. Throughout
the paper, we use natural units where 
$c = \hbar = k_B = 1$.

\section{Symmetric Dark Atom Scenario}

In the dark atom scenario, the dark sector contains
a heavier fermion $\chi_p$  with mass $m_{\chi_p}$
and its lighter partner $\chi_e$ with mass $m_{\chi_e}$, 
as well as their antiparticles $\bar \chi_p$, $\bar \chi_e$.
$\chi_p$ and $\chi_e$ interact through a long-range force with opposite charges, which allows the formation of a bound state. 
Throughout this paper, we take the long-established dark 
$U(1)_X$ as an example for implementation.
The bound atomic state 
$\chi_A \equiv (\chi_p \chi_e)$ shall form when the temperature is below their binding energy $E_b$.
The subscripts $p$ and $e$ are analogies to the proton and electron in the SM. The dark atom $\chi_A$ and its antiparticle account for the DM today.

Our main body Lagrangian is the same as the usual dark atom model~\cite{Kaplan:2009de,Kaplan:2011yj}. However, we do not have the additional assumption of atom-antiatom asymmetry. The Lagrangian is
\begin{align}
  \mathcal L
& \supset
  \frac 1 4  \epsilon F^{\mu \nu} F'_{\mu \nu}
- \frac 1 4  F^{\mu\nu} F_{\mu\nu}
- \frac 1 4  F^{\prime \mu\nu} F_{\mu\nu}'
+ \frac12 m_{A'}^2 A'^\mu A'_\mu 
\notag
\\
&\quad +
\bar{\chi}_p (i\slashed{D} -m_{\chi_p} )\chi_p 
+ \bar{\chi}_e (i\slashed{D} - m_{\chi_e} )\chi_e.
\label{eq:minimalL}
\end{align}
Here, $A'$ and $F'$ are the dark $U(1)_X$ gauge boson and field strength. 
As usual, the dark proton $\chi_p$ carries $U(1)_X$ charge $+1$ while dark electron $\chi_e$ has charge $-1$. 
This dark $U(1)_X$ has a mixing with the SM $U(1)_{\rm em}$ by 
a mixing angle $\epsilon$.
The dark gauge boson has a tiny 
mass to mediate a long-range force and exists as dark radiation.
With a 
mixing satisfying the current limit $\epsilon \lesssim 10^{-12}$~\cite{Schwarz:2015lqa,Caputo:2021eaa}, the dark sector cannot 
always be in kinetic equilibrium with 
the SM sector and will attain its own temperature $T_\chi$. To avoid the 
big bang nucleosynthesis (BBN) constraint for $A'$, we take 
$T_\chi = T_{\rm SM} \xi$ and $\xi = 0.2$~\footnote{ Roughly the extra effective relativistic degree of freedom should be smaller than 0.28~\cite{ParticleDataGroup:2022pth} measured from CMB. This translates to a constraint on the temperature ratio, which is $T_\chi<0.5 \, T_{\rm SM}$~\cite{Cline:2021itd}.}.
Both the dark proton and dark electron maintain in 
chemical equilibrium inside the dark sector through their annihilation into 
gauge bosons, $\bar \chi_{p(e)} + \chi_{p(e)} \leftrightarrow 2 A'$ 
in the very beginning.
The thermal averaged cross section $\braket{\sigma^{p(e)}_{\rm ann} v} \simeq \alpha_{D}^2/m^2_{\chi_{p(e)}}$
is proportional to the square of particle wavelength.
Since the dark proton mass is 
much larger than that of electron and binding 
energy, $\chi_p$ will first freeze out through
such a channel and its yield $Y_{\chi_p}=n_{\chi_p}/s$ stays as a constant for the moment.

As long as the mass of dark $U(1)_X$ gauge boson $A'$ is much smaller than both fermion masses, we can treat the interaction between dark fermions as a Coulomb potential~\cite{Napsuciale:2020ehf}, and describe dark atoms
by a simple Bohr model with high accuracy.
The binding energy and Bohr radius of the dark atom are
\begin{equation}
  E_b 
= 
  \frac12 \alpha_{D}^2 \mu
\,,
\quad
r_b = \frac{1}{\alpha_{D} \mu}\;,
\end{equation}   
where dark fine structure constant $\alpha_{D} \equiv g^2/4\pi$ and reduced mass $\mu \equiv (m_{\chi_p} m_{\chi_e})/(m_{\chi_p} + m_{\chi_e}) \simeq m_{\chi_e}$. With $\alpha_{D} < 1$, 
the dark atom size is larger than the Compton wavelength of both $\chi_p$ and $\chi_e$.

The dark sector temperature is roughly the kinetic energy of particles inside it. 
When the Universe cools down to $T_\chi \sim E_b$, the relative kinetic energy between 
dark protons and electrons becomes 
smaller than the atomic binding energy. After that, both atom and antiatom begin to form.
The atomic formation (AF) cross section is~\cite{sobel2016introduction},
\begin{equation}
    \langle \sigma_{\rm AF} v \rangle
= 
  \frac{16 \pi}{3 \sqrt 3}
  \frac{\alpha_D^2}{\mu^2}
  \left(\frac{E_b}{T_\chi} \right)^{1 / 2} 
  \ln \left(\frac{E_b}{T_\chi} \right).
\label{AtmoicFormation}
\end{equation}

Symmetry between atom and antiatom indicates that their
annihilation also happens once number density accumulates.
In the nonrelativistic limit, the annihilation is dominated by the 
atomic rearrangement (AR) processes~\cite{Morgan:1973zz,froelich2000hydrogen}, 
\begin{subequations}
\begin{align}
    (\chi_p \chi_e) + 
    (\bar \chi_p \bar \chi_e)
    & \rightarrow
    (\bar \chi_p \chi_p)
    + \bar \chi_e + \chi_e\;,
\label{eq:ar_1}\\
   (\chi_p \chi_e) + 
    (\bar \chi_p \bar \chi_e)
    &\rightarrow
    (\bar \chi_p \chi_p) + 
    (\bar \chi_e \chi_e)\;,
\label{eq:ar_2} %
\end{align}%
\label{eq:ar}%
\end{subequations}%
The bound states $(\bar\chi_p \chi_p)$ and
$(\bar\chi_e \chi_e)$ shall decay to dark photons subsequently once they are formed.
Thus, the atomic annihilation is effectively one directional.
In principle, $\chi_p$ and $\bar \chi_p$ can directly annihilate in flight without forming the intermediate $(\bar{\chi}_p \chi_p)$, but
the cross section is usually small compared to the rearrangements~\cite{froelich2000hydrogen}.
The above processes have a geometrical cross section~\cite{Morgan:1973zz,froelich2000hydrogen},
\begin{equation}
  \langle \sigma_{\rm AR} v \rangle
\simeq 
  \mathcal{C} \pi r_b^2.
\end{equation}
The numerical prefactor $\mathcal{C}\sim\mathcal{O}(1)$ could be calculated by investigating the potential between bound 
states. Generally,
this factor depends on the collision energy. 
However, in low-energy regions, this dependence is quite
weak~\cite{froelich2000hydrogen}. 
So, we treat it as constant and take $\mathcal C = 1$ for both processes.~\footnote{Since cross sections for processes in \eqref{eq:ar_1} and \eqref{eq:ar_2} are both geometrical, they are naturally of the same order. Here we assume that $\mathcal{C}=1$ for simplicity.}
The $\langle \sigma_{\rm AR} v\rangle$ is roughly the geometrical size of the atom.
As shown in~\gfig{fig:CrossSection},
before and after the formation of the bound state, 
the DM size increases by a factor of $\alpha_{D}^{-4} (m_{\chi_p}/m_{\chi_e})^2$, which 
significantly enhance the annihilation cross section if $m_{\chi_p} \gg m_{\chi_e}$.
The DM annihilation could happen again even with an already depleted density. 
This further decreases the dark fermion number, and the freeze-out of symmetric dark atoms is then 
finally determined by atomic rearrangement. As we will see later, 
the DM mass is ultraheavy, even beyond the unitarity 
bound, to give the observed relic.

Apart from the processes in Eq.~\eqref{eq:ar}, the reaction $\chi_p + (\bar \chi_p \bar\chi_e) \rightarrow (\bar\chi_p \chi_p) + \bar \chi_e$ and its conjugate reaction occur simultaneously. Their cross sections, denoted as $\sigma_{p\bar{A}}$, are also geometric in nature, with $\sigma_{p\bar{A}} v \simeq \pi r_b^2$.
On the other hand, the rearrangement of $\chi_e$ and the dark atom is kinetically suppressed at low energy. This is because the binding energy of $(\chi_e \bar{\chi}_e)$ is smaller than that of the dark atom, making the reaction endothermic.
As we will discuss later, while the rearrangement annihilation between dark protons and dark atoms leads to a rapid depletion of dark protons once a certain number of dark atoms is produced, the ultimate freeze-out of the dark atom is still primarily determined by the processes described in Eq.~\eqref{eq:ar}.

However, the minimal thermal symmetric dark atom model in Eq.~\eqref{eq:minimalL} has an intrinsic problem.
To efficiently consume the dark proton number and convert them into atoms, its lighter 
partner $\chi_e$ should have almost the same 
yield at the time $T_\chi \sim E_b$.
The cross section of dark fermion annihilation into the dark boson is inversely related to its mass square. 
It means that $\chi_e$ has a larger annihilation rate and stays in equilibrium for a longer time. The dark electron 
has a smaller yield than that of 
the dark proton when it freezes out.
As a result, only a small part of the dark proton can be consumed to form atoms while others are left as millicharged particles.
In such case, the atomic annihilation effect becomes negligible and the heavy dark fermion density shall exceed the total DM density. 

One way to solve this problem is to use a real scalar particle $\phi$, which can be the real component of dark Higgs from UV completion of $U(1)_X$, with mass $2 m_e < m_\phi < 2 m_p$. It couples to dark fermions through the Yukawa interaction,~\footnote{In principle, this scalar can couple to SM Higgs by $\lambda_{1} \Lambda \phi H^\dagger H +  \lambda_{2} \phi^2 H^\dagger H$.
However, the interaction should be small to  
avoid possible constraints from Higgs interaction.}
\begin{equation}
\mathcal{L}
\supset
    y_p \phi \bar{\chi}_p \chi_p
    +
    y_e \phi \bar{\chi}_e \chi_e\;.
\end{equation}
Both Yukawa couplings, $y_p$ and $y_e$, are small, so that 
the scalar $\phi$ does not come into either kinetic or chemical equilibrium with the dark sector. 
Its merit is to continuously 
decay to $\chi_e$ to help the atomic formation 
process. 

With the model and settings, we are ready 
to discuss the whole thermal freeze-out history of 
our symmetric atomic DM scenario.

\section{Freeze-Out through Atomic Rearrangement}

\begin{figure}[t]
    \centering
     \includegraphics[width=8cm]{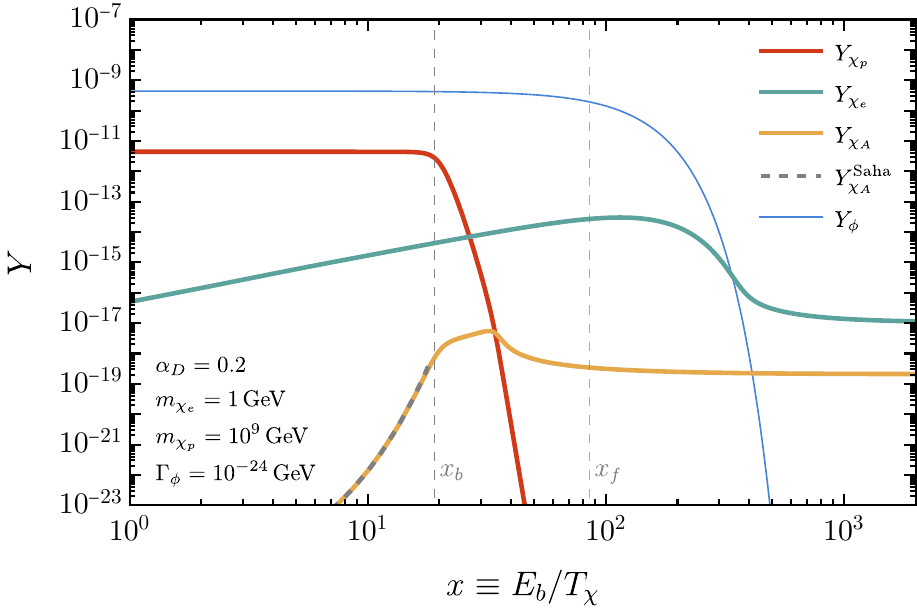}
    \caption{
    The yield evolution of dark proton $\chi_p$ (red), dark electron $\chi_e$ (green), scalar $\phi$ (blue), and dark atom $\chi_A$ (yellow). 
    We take $m_{\chi_e} = 1\,$GeV, 
    $\alpha_D = 0.2$, $\Gamma_\phi = 10^{-24}\,$GeV, 
    and $Y_{\phi}^0 = 100 Y_{\chi_p}^0$.
    The yield of dark atom initially increases due to its formation, then decreases as a result of its rearrangement, and finally freezes out.
    To achieve the observed relic 
    $\Omega_{\chi_A} h^2 = 0.12$, the dark 
    proton mass is $m_{\chi_p} = 10^{9}\,$GeV.
    For comparison, the atom yield solution of the Saha equation is drawn as a gray dashed line. }
    \label{fig:yield}
\end{figure}

This section shows our scenario
naturally introduces a new freeze-out mechanism driven by atomic rearrangement.
In the early Universe, the symmetric dark atom 
scenario has three typical energy scales: the
dark proton mass $m_{\chi_p}$, the dark electron mass $m_{\chi_e}$, and binding energy $E_b$.
The unitarity limit of cross section 
requires 
the dark coupling to be $\alpha_D \leq 0.5$ \cite{vonHarling:2014kha}.
In this paper, the 
dark proton mass is ultraheavy and the mass difference between fermions
is huge.
Thus, there is a hierarchy 
$m_{\chi_p} \gg m_{\chi_e} \gg E_b$. 
The whole freeze-out histories of $\chi_p$, $\chi_e$, $\phi$, and $\chi_A$
have the following clear phases
as shown in \gfig{fig:yield}. Here, the temperature parameter is defined by
$x \equiv E_b/T_\chi$ and 
the parameters are taken as
$\alpha_D = 0.2$, $m_{\chi_e} = 1\,$GeV,
and $\Gamma_\phi = 10^{-24}\,$GeV.

\subsection{$\chi_p$ Freeze-out ($T_\chi \simeq \mathcal{O}(m_{\chi_p})$)} 
In this phase, the temperature is so high that all other species remain relativistic and in equilibrium except $\chi_p$.
When the dark sector temperature is below $m_{\chi_p}$, the dark proton 
becomes nonrelativistic and 
begins to freeze out through the channel $\chi_p + \bar \chi_p \leftrightarrow 2 A'$.
For large $U(1)_X$ coupling $\alpha_{D}$,
the Sommerfeld enhancement and the bound-state formation effects should be included 
in the annihilation cross section, $\left\langle \sigma^p_{\rm ann} v \right\rangle = (\alpha_{D}^2/m_{\chi_p}^2) \times \mathcal{S}$, 
as an effective enhancement factor $\mathcal S$~\cite{vonHarling:2014kha}.
The dark proton yield after this first stage freeze-out $Y_{\chi_p}^0 $ can be fixed by its 
mass, coupling, and the enhancement factor. 
Since the $\chi_p$ mass is much larger than that of $\chi_e$, this first stage freeze-out temperature $m_{\chi_p}/20$ is far above binding energy $E_b$.
Thus, no dark atom is formed and the yield $Y_{\chi_p}^0$ stays as a constant to give the initial value of the red curve until $T_\chi \sim E_b$. 
Currently, $\chi_p$ with a mass larger than $\mathcal{O}(100)$\,TeV is overproduced \cite{Griest:1989wd},
and the consumption by atomic 
rearrangement in later phases 
is necessary.

At the same time, the scalar $\phi$
gradually freezes in through the channel, 
$\bar \chi_p + \chi_p \rightarrow \phi + A'$.
Its yield after freeze-in
$Y_\phi^0$
is proportional to the 
annihilation cross section 
of order 
$y_p^2 \alpha_D$. It also 
depends on how long it lasts,
$ 1/T_{fi}$, where the $T_{fi}
\sim \mathcal{O}(m_{\chi_p})$
is the endpoint of freeze-in.
By integrating the Boltzmann equation of $\phi$,
one can get,
\begin{equation}
    Y^0_\phi \simeq 
    4 \times 10^{-7}
    \frac{y_p^2 \alpha_D M_P }{g_*^{3/2} T_{fi}}.
\label{Yieldphi}
\end{equation}

\subsection{Atomic Formation ($T_\chi \gtrsim E_b / x_b \simeq E_b / 30$)} 

When the dark sector temperature cools down to $T_\chi \sim E_b$, the dark electrons begin to pair
with the dark protons and form atoms.
The density evolution of the dark proton $\chi_p$, dark electron $\chi_e$, and atom state $\chi_A$ are described by the following Boltzmann equations with respect to 
time $t$:
\begin{subequations}
\begin{align}
  \frac{d Y_{\chi_p}}{d t}
& = 
-  s
  \langle \sigma_{\rm AF} v \rangle 
  \left( Y_{\chi_p} Y_{\chi_e} - Y_{\chi_p}^{\mathrm{eq}} Y_{\chi_e}^{\mathrm{eq}} \frac{Y_{\chi_A}}{Y_{\chi_A}^{\mathrm{eq}}} \right) \notag \\
  & 
  \quad - s
  \langle \sigma_{p \bar{A}} v \rangle 
  Y_{\chi_p} Y_{\chi_A}
  \;, 
\label{chip:stage2}
\\
  \frac{dY_{\chi_e}}{dt}
&= 
    - s \langle \sigma^e_{\rm ann} v \rangle 
    \left( Y^2_{\chi_e} - (Y_{\chi_e}^{\mathrm{eq}})^2  \right) \notag
\\
& \quad 
- 
  s \langle \sigma_{\rm AF} v \rangle 
  \left( Y_{\chi_p} Y_{\chi_e} - Y_{\chi_p}^{\mathrm{eq}} Y_{\chi_e}^{\mathrm{eq}} \frac{Y_{\chi_A}}{Y_{\chi_A}^{\mathrm{eq}}} \right) \label{Beq:Xe}
\\
& \quad +
  \left\langle \Gamma_\phi \right\rangle
  Y_\phi
  +
  s \left\langle \sigma_\mathrm{AR} v \right\rangle 
  Y_{\chi_A}^2
  + s
  \langle \sigma_{p \bar{A}} v \rangle 
  Y_{\chi_p} Y_{\chi_A}
  \;,\notag
\\
  \frac{d Y_{\chi_A}}{dt}
&= 
  s \left\langle \sigma_\mathrm{AF} v \right\rangle 
  \left( Y_{\chi_p} Y_{\chi_e} - Y_{\chi_p}^{\mathrm{eq}} Y_{\chi_p}^{\mathrm{eq}} \frac{Y_{\chi_A}}{Y_{\chi_A}^{\mathrm{eq}}} \right) \notag
\\
& \quad - 
  2 s \langle \sigma_\mathrm{AR} v \rangle 
  Y_{\chi_A}^2
  - s
  \langle \sigma_{p \bar{A}} v \rangle 
  Y_{\chi_p} Y_{\chi_A}
  \;.
\label{Beq:A}
\end{align}
\end{subequations}
The initial conditions are 
$\left. Y_{\chi_p (\phi)} \right|_{T_\chi = E_b} = 
Y_{\chi_p (\phi)}^0$, $Y_{\chi_A}^0|_{T_\chi = E_b} = 0$.
Notice that the inverse 
process of the atomic 
rearrangement is neglected.
Here, $Y_i^{\rm eq} = n_i^{\rm eq}/s$ and $n_i^{\rm eq}$ is the number density in equilibrium for the corresponding species. 
The factor $2$ in the second term on the right of \geqn{Beq:A} comes from the inclusion of two rearrangement processes~\geqn{eq:ar}.
The evolution of antiparticles is the same.

The dark electron 
$\chi_e$ is injected into the Universe all along this phase to pair with dark protons
as the green curve.
The yield of $\chi_e$ is determined by the 
scalar $\phi$ decay, $\phi \rightarrow \bar \chi_e + \chi_e$,
and its own annihilation to dark gauge boson including Sommerfeld and bound state formation enhancements.
To make sure the $\chi_e$ number is large
enough, the initial yield of $\phi$ in \geqn{Yieldphi} should be larger than 
the yield of $\chi_p$ as $Y_{\phi}^0 \geq Y_{\chi_p}^0$. 

On the other hand,
dark atoms are constantly being formed and dissociated to maintain in equilibrium. 
Its evolution can be analytically solved through the Saha equation~\cite{saha1920liii,saha1921physical} [\geqn{Beq:A} with only atomic formation term
since $Y_{\chi_A}$ is extremely small compared to $Y_{\chi_p}$ in this stage]. 
The solution, 
    \begin{equation}
        n_{\chi_A}^{\mathrm{Saha}} = 
        \frac{n_{\chi_p} n_{\chi_e} }{n_{\chi_p}^{\mathrm{eq}} n_{\chi_e}^{\mathrm{eq}} }
        n_{\chi_A}^{\mathrm{eq}} \;,
        \label{saha_eq}
    \end{equation}
shown as the gray dashed line, perfectly 
fits the yield evolution of the atom (yellow curve). At this stage, the
$Y_{\chi_p}$ (red curve) remains constant 
while the number density of dark atoms 
increases exponentially, $n_{\chi_A}^{\rm Saha} 
\propto e^{E_b /T}$.

\subsection{Atomic Rearrangement ($T_\chi \gtrsim E_b / x_f \simeq E_b / 100$)}

As $n_{\chi_A}$ accumulates, at the temperature $T_\chi = T_b$ (with $x_b \equiv E_b/T_b$),
the rearrangement annihilation of dark proton and dark atoms dominates over the Hubble dilution of $n_{\chi_p}$,
\begin{equation}
        n_{\chi_A} \braket{\sigma_{p \bar{A}} v} 
        > 
        H \,,
        \label{rearragement_condition}
    \end{equation}
where $H (T_{\rm SM}) = 2 \sqrt{\pi^3 g_*/45} T_{\rm SM}^2/M_P$ with the Planck mass $M_P \equiv 1 / \sqrt{G_N} \approx 1.2 \times 10^{19}~{\rm GeV}$.
Meanwhile, the atomic formation process is also intense:
\begin{equation}
    n_{\chi_e} \braket{\sigma_{\rm AF} v} > H\;.
    \label{e_condition}
\end{equation}
Dark protons, together with dark electrons, quickly form dark atoms and annihilate with antiatoms through atomic rearrangement. So the yield of $\chi_p$
drops sharply right after $x_b$ as shown in the red curve.

Shown as the yellow curve, the atomic annihilation also decreases the yield of atom $Y_{\chi_A}$. 
It becomes more and more difficult for dark atoms to find each other and rearrange. 
Atomic rearrangement freezes out at the temperature $T_\chi = T_f$ (with $x_f \equiv E_b/T_f$) and Hubble dilution takes over. 
The eventual freeze-out point is determined by,
    \begin{equation}
        n_{\chi_A} \braket{\sigma_{\rm AR} v} \simeq H\;.
        \label{freeze-out-condition}
    \end{equation}
The relic of dark atoms makes up the observed DM today.
The final dark atom yield $Y^f_{\chi_A}$ is predicted as,
\begin{equation}
    Y_{\chi_A}^f \simeq \frac{3\sqrt{5}}{\pi^{3/2} \sqrt{g_*}} \frac{m_{\chi_e} x_f \xi}{M_P} \;.
\end{equation}
To fit the observed relic
$\Omega_{\chi_A} h^2 
    = 
    2 m_{\chi_A} s_0 Y^f_{\chi_A} h^2 / \rho_c = 0.12$, where $s_0 (\rho_c)$ denotes the entropy (critical) density today, the DM mass should be
\begin{equation}
\hspace{-1mm}
     m_{\chi_A} 
 \simeq 
    10^{9}
    \,{\rm GeV}
\left( \frac{1~{\rm GeV}}{m_{\chi_e}} \right)
    \left( 
    \frac{g_*}{10}
    \right)^{1/2}
    \left( \frac{80}{x_f} \right)
    \left( \frac{0.2}{\xi} \right).
\label{massDependence}
\end{equation}
The final yield and DM mass depend simply on the parameter $m_{\chi_e}$ and are not
sensitive to initial yields before 
atomic formation. Even the dependence on coupling 
$\alpha_D$ is canceled out.
Besides,  
the final yield of the atom 
$Y_{A}^f$ decreases a lot
compared with $Y_{\chi_p}^0$.
Thus, the dark proton mass 
can be lifted to be ultraheavy as
$10^9\,$GeV in the case of 
\gfig{fig:yield}.

After all the phases are finished, 
a large part of $\phi$ begins to decay, $\Gamma_\phi \simeq y_e^2 m_\phi / 4\pi \simeq H(E_b/x_f \xi)$. Without $\phi$ injection, the dark electron will finally annihilate
and freeze out.
Since the mass difference is huge,
$m_{\chi_e} \ll m_{\chi_p} \approx m_{\chi_A}$, the millicharged
dark electron only accounts for a tiny portion of DM.

So far, we have seen the key points of the symmetric dark atom scenario.
Its freeze-out is determined by atomic rearrangement and can give the observed 
relic. 
Since the geometric cross section of atom annihilation is much larger than the annihilation cross section of $\chi_p$ itself as a pointlike particle, 
the number density of dark sector particles is greatly suppressed, and its mass is enhanced to be ultraheavy.

\begin{figure}[t]
    \centering
     \includegraphics[width=8cm]{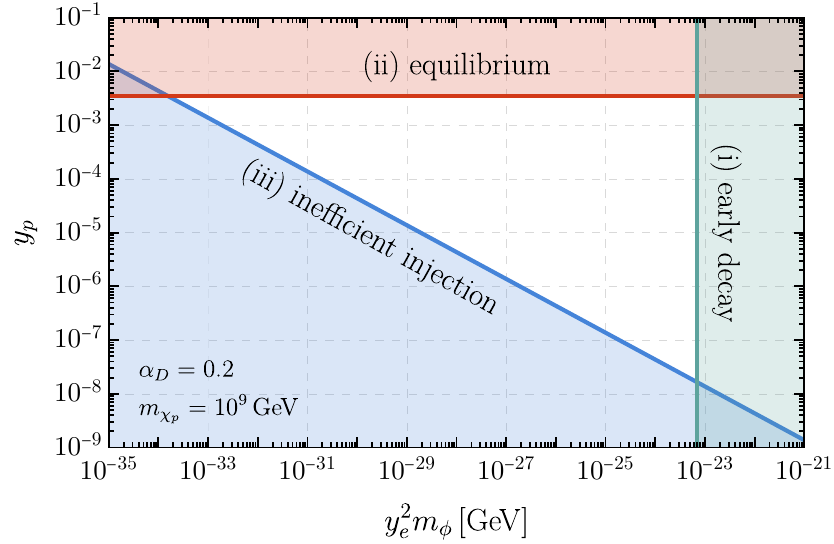}
    \caption{
    The parameter space for $\phi$ in the $(y_e^2 m_\phi, y_p)$ plane that successfully initiates the thermal freeze-out of the symmetric dark atom.
    Here we take $\alpha_D = 0.2$ and $m_{\chi_p} = 10^9 \, \mathrm{GeV}$.
    }%
    \label{fig:phi_para_space}
\end{figure}

\section{Parameter Space and Constraints}

We first explore the required parameter space of $\phi$ to make the symmetric atomic DM scenario work.
(i) We expect the decay of $\phi$ to happen after the freeze-out of the dark atom, $\Gamma_{\phi} < H(x_f)$. 
Otherwise, the injection of $\chi_e$ during atomic formation is not sufficient and $\chi_p$ cannot be significantly depleted.
(ii) Since $\phi$ is produced by freeze-in and never enters equilibrium, its yield after freeze-in should be smaller than its relativistic equilibrium yield, $Y_{\phi}^0 < T_\chi^3/\pi^2 s \simeq 0.0018 g_*^{-1}$.
(iii) The yield of $\chi_e$ before $x_b$ is controlled by two major processes, its self-annihilation and the decay of $\phi$. 
The depletion of $\chi_e$ due to its self-annihilation should be compensated for by the production from the decay of $\phi$.
Its yield is approximately $Y_{\chi e} \approx \sqrt{\Gamma_\phi Y_{\phi}^0/s \langle \sigma^e_{\rm ann} v \rangle}$. 
To meet the criterion of the minimal number density of $\chi_e$ 
as Eq.~\eqref{e_condition}, 
one obtains another requirement for $\phi$, which is $\Gamma_\phi Y_{\phi}^0 > H(x_b)^2  \langle \sigma^e_{\rm ann} v \rangle / s \left\langle \sigma_\mathrm{AF} v \right\rangle^2 $.
In Fig.~\ref{fig:phi_para_space}, we show the allowed parameter space (blank area) that satisfies these three conditions for $\phi$ by taking 
$\alpha_D = 0.2$ and $m_{\chi_p} = 10^9\,$GeV.

\begin{figure}[t]
    \centering
     \includegraphics[width=8cm]{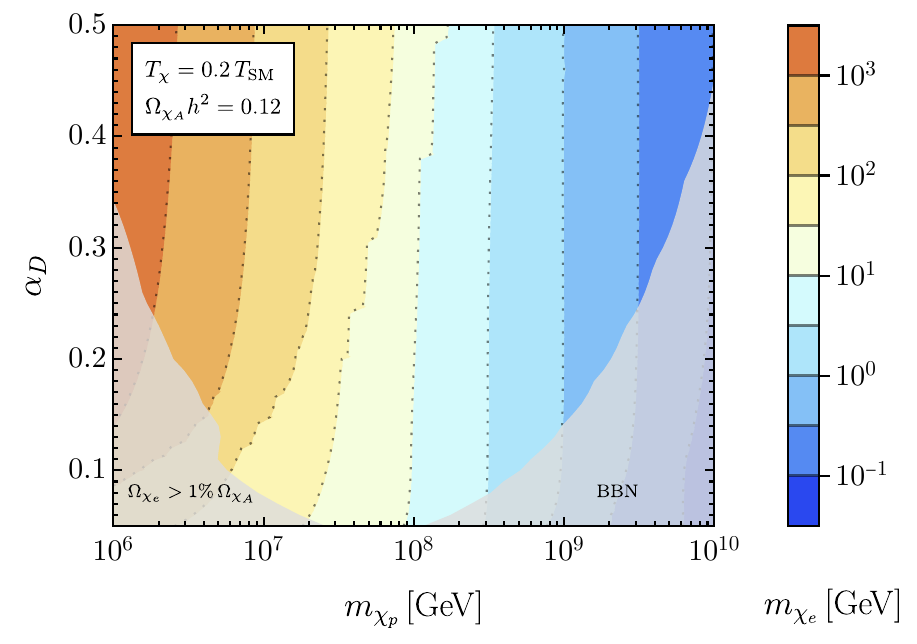}
    \caption{
   The parameter space where dark atom is giving the 
   observed DM relic $\Omega_{\chi_A} h^2 = 0.12$.
   The horizontal axis is the DM mass
   $m_{\chi_A} \approx m_{\chi_p}$,
   and the vertical axis is the coupling 
   $\alpha_D$. Different colors stand for
   different values of dark electron mass
   $m_{\chi_e}$. The DM mass is not sensitive to the coupling but inversely proportional to dark electron mass.
   The gray shaded area 
   is excluded by the BBN constraint
   and the overproduction of 
   $\chi_e$,
   $
   \Omega_{\chi_e}
   >
   1\% \,\Omega_{\chi_A} 
   $.}
    \label{fig:parameter}
\end{figure}

The parameter space of the symmetric atomic DM to give the correct DM relic $\Omega_{\chi_A} h^2 = 0.12$ is shown in \gfig{fig:parameter}.
Here, we fix the scalar yield after its 
freeze-in as $Y^0_{\phi} = 100\times Y_{\chi_p}^0$,
and it decays around 
$x = E_b / T_\chi \sim 100$
by choosing 
appropriate $y_p$, $y_e$, and $m_\phi$.
Once the above conditions are met, the impact of different parameter values of $\phi$ on the dark atom abundance becomes negligible. 
The DM relic then depends on three parameters,
dark fermion masses $m_{\chi_p}$, $m_{\chi_e}$ and the dark $U(1)_X$ coupling $\alpha_D$.
One can see that the DM mass $m_{\chi_A} \approx m_{\chi_p}$ can be in the range of $(10^6, 10^{10})\,$GeV by varying $\alpha \subset (0.05, 0.5)$ and $m_{\chi_e} \subset (10^{-1}, 10^3)\,$GeV.
Exactly as expected in 
\geqn{massDependence}, the 
DM mass $m_{\chi_A} (m_{\chi_p})$ is
not sensitive to coupling $\alpha_D$
but inversely proportional to the dark electron mass $m_{\chi_e}$.
The reason is as follows.
A lighter dark electron leads to a larger atomic rearrangement cross section.
The dark atom yield becomes smaller after its freeze-out.
As a result, the DM shall be heavier 
to give the correct relic.

The parameters in the bottom-left corner of the figure lead to an  overproduction 
of $\chi_e$ by the criterion
$\Omega_{\chi_e} > 1\% \, \Omega_{\chi_A}$.
For a larger $m_{\chi_e}$ 
and a smaller $\alpha_D$, the annihilation cross section of dark 
electron is suppressed. 
Thus, more dark electrons are left
as relic. 
The region where the dark fermions have a small mass difference is disfavored.
It indicates that the freeze-out through atomic rearrangement naturally produces DM heavier than the unitarity bound.

If the binding energy $E_b \propto
\alpha_D^2 m_e$ is too small, the atomic formation and annihilation happen around the BBN epoch. The number and energy density 
of $\chi_p$ before consumption can be large enough to affect BBN. 
Thus, the bottom-right corner is 
excluded as shown by the 
gray shaded area.

\section{Conclusion and Discussion}
In this paper, we propose a new scenario in which DM is composed of both dark atoms and antiatoms symmetrically and discuss its freeze-out production.
In this scenario, a heavier dark fermion $\chi_p$ pairs with a lighter partner $\chi_e$ to form an atomic state.
Since the cross section of atomic annihilation through rearrangement is much larger than the unitarity limit of $\chi_p$ annihilation,
the dark sector particles are 
further consumed after atomic formation. The dark atom freeze-out
is determined by their rearrangements.
Notably, the mass of DM can avoid the unitarity bound and be lifted to $\mathcal{O}(10^{10})\,$GeV.

About the possible 
signatures of the symmetric dark atom scenario,
we have the following open discussions.
Compared with the asymmetric 
dark atom scenario, the symmetric case 
has different phenomena for the annihilation between particle and antiparticle
appears.
This annihilation can be tested through 
indirect detection and cosmological observation.
Usually, such constraints are not sensitive to heavy DM because of the small number density~\cite{Elor:2015bho,Kawasaki:2021etm}. 
However, the large annihilation cross section of the symmetric DM is a remedy for this deficiency.  
The ultraheavy symmetric atomic DM is self-interacting with a huge cross section. 
Thus, its halo density profile could be different from those of other models. This difference also imprints in the matter power spectrum at small scales.
Furthermore, the annihilation of atomic DM shall distort cosmic microwave background (CMB) and leave indirect detection signals today.
These phenomenological research studies will be explored in future works.

\section*{Acknowledgements}

The authors thank Yu Cheng and Chen Xia for useful discussions. 
Jie Sheng deeply thanks Professor Shao-Feng Ge, Tsutomu Yanagida san, and 
Mengfan Shi for
their encouragement.
This work is supported by the National Natural Science Foundation of China
(12247141, 12247148, 12375101, 12090060, 12090064)
and the SJTU Double First Class start-up fund (WF220442604).

\bibliography{DarkAtom}

\end{document}